\documentclass{aa}
\usepackage{graphicx}
\usepackage{natbib}
\usepackage{txfonts}
\bibpunct{(}{)}{;}{a}{}{,} 

\def\ls{LS~5039}
\def\lsi{LSI~+61$\degr$303}
\def\psrb{PSR~B1259-63}
\begin{document}

\title{The modulation of the gamma-ray emission from the binary LS~5039}
\titlerunning{Gamma-ray modulation from the binary LS 5039}

\author{
Guillaume Dubus 
\and Beno\^it Cerutti 
\and Gilles Henri 
}
\authorrunning{Dubus, Cerutti and Henri}
\institute{
Laboratoire d'Astrophysique de Grenoble, UMR 5571 CNRS, Universit\'e
Joseph Fourier, BP 53, 38041 Grenoble, France 
}

\date{Draft \today}
\abstract
{Gamma-ray binaries have been established as a new class of sources of very high energy (VHE, $>$100~GeV) photons. These binaries are composed of a massive star and a compact object. The gamma-rays are probably produced by inverse Compton scattering of the stellar light by VHE electrons accelerated in the vicinity of the compact object. The VHE emission from LS~5039 displays an orbital modulation. }
{The inverse Compton spectrum depends on the angle between the incoming and outgoing photon in the rest frame of the electron. Since the angle at which an observer sees the star and electrons changes with the orbit, a phase dependence of the spectrum is expected.}
{A procedure to compute anisotropic inverse Compton emission is
  explained and applied to the case of \ls. The spectrum is calculated
  assuming the continuous injection of electrons close to the compact
  object: the shape of the steady-state distribution depends on the
  injected power-law and on the magnetic field intensity.}
{Compared to the isotropic approximation, anisotropic scattering
  produces harder and fainter emission at inferior conjunction,
  crucially at a time when attenuation due to pair production of the
  VHE gamma-rays on star light is minimum. The computed lightcurve and
  spectra are very good fits to the HESS and EGRET observations,
  except at phases of maximum attenuation where pair cascade emission
  may be significant for HESS. Detailed predictions are made for a
  modulation in the GLAST energy range. The magnetic field intensity
  at periastron is 0.8$\pm0.2$~G.
 }
{The anisotropy in inverse Compton scattering plays a major role in
  \ls.  A simple model reproduces the observations, constraining the magnetic field intensity and injection
  spectrum.  The comparison with observations, the derived magnetic field
  intensity, injection energy and slope
suggest emission from a rotation-powered pulsar wind nebula. These results confirm gamma-ray binaries as promising
  sources to study the environment of pulsars on small scales.}
\maketitle

\section{Introduction}
Gamma-ray binaries have been established in the past couple of years
as a new class of sources of very high energy (VHE, $>$100~GeV)
photons. They are characterized by a large gamma-ray luminosity above
an MeV, at the level of or exceeding their X-ray luminosity. At
present, all three such systems known (\object{\ls}, \object{\psrb}\
and \object{\lsi}, recently possibly joined by \object{Cyg X-1})
comprise a massive star
\citep{2005Sci...309..746A,2005A&A...442....1A,2006Sci...312.1771A,2007arXiv0706.1505M}.
The compact object in \psrb\ is a 48-ms, young radio pulsar. The VHE
emission arises from the interaction of the relativistic wind from
this pulsar, extracting rotational energy from the neutron star, with
the stellar wind from its companion \citep{1994ApJ...433L..37T}.
Particles gain energy at the shock between the winds, resulting in a
small-scale pulsar wind nebula \citep{1981MNRAS.194P...1M}. The
particles radiate away their energy as they are entrained in the
shocked flow, forming a comet-like trail of emission behind the pulsar
\citep{2006A&A...456..801D}.

The nature of the compact object and origin of the VHE emission
remains controversial in \ls\ and \lsi, although recent observations
indicate the radio emission of \lsi\ behaves like the comet tail
expected in the pulsar scenario \citep{2006smqw.confE..52D}.
Alternatively, the VHE emission could originate from particles
accelerated in a relativistic jet, the energy source being accretion
onto a black hole or neutron star
\citep{2006ApJ...643.1081D,2006A&A...451..259P}. The rationale being
that there is evidence for particle acceleration in the jets of
microquasars and active galactic nuclei.  However, hard evidence for
accretion occuring in either \ls\ or \lsi\ has been hard to come by
\citep[e.g.][]{2005A&A...430..245M} and the similarities between the
three systems (and differences with the usual microquasars) do not
argue in favour of the accretion/ejection scenario
\citep{2006A&A...456..801D}.

Regardless of the actual powering mechanism, some particles must be
accelerated to high energies to generate the VHE gamma-rays. If these
particles are leptons, the only viable gamma-ray radiation mechanism
is inverse Compton scattering on the stellar photons.  The massive
stars in gamma-ray binaries have effective temperatures of several
tens of thousand K and radii of about 10~$R_\odot$, yielding
luminosities of the order of $10^{39}$~erg~s$^{-1}$. This provides a
huge density of stellar photons in the UV band that VHE leptons
may up-scatter, much greater than any other possible source of target
photons (e.g. synchrotron or bremsstrahlung emission).

The emitted VHE photons also have enough energy to produce $e^+e^-$
pairs with the UV stellar photons. Most of the VHE flux may therefore
be lost to the observer if the source is behind the star and VHE
photons have to travel through the stellar light. Gamma-ray
attenuation has been shown to lead to a modulation of the VHE flux
with minimum absorption (maximum) at inferior (superior) conjunction
\citep{2005ApJ...634L..81B,2006A&A...451....9D}.

HESS observations have indeed shown a stable modulation of the VHE
flux from \ls\ on the orbital period with a maximum around inferior
conjunction. This suggests attenuation plays a role and that the
source of VHE gamma-rays cannot be more than about an AU from the
binary (or attenuation would be too weak to modulate the flux).
However, a non-zero flux is detected at superior conjunction where a
large attenuation is expected, possibly because of pair cascading.
Moreover, the spectral changes that are reported do not fit with an
interpretation based on pure attenuation of a constant VHE source
spectrum \citep{2006A&A...460..743A}.

Inverse Compton scattering also has a well-known dependence on the
angle $\Theta$ between incoming and outgoing photon.  The photon flux
from the star being anisotropic, the resulting inverse Compton
emission will depend on the angle at which it is observed. Hence, a
phase-dependent VHE spectrum will be observed even if the distribution
of particles is isotropic and remains constant throughout. This effect
has previously been investigated in \psrb\ by
\citet{2000APh....12..335B} who calculated the radiative drag on the
unshocked pulsar wind from scattering of stellar light, using results
from \citet{1989ApJ...343..277H}.  The drag produces a Compton
gamma-ray line with a strong dependence on viewing angle. 

This work purports to explain the HESS observations of LS~5039
using a combination of anisotropic inverse Compton scattering and
attenuation in the simplest way possible. The aim is to
constrain the underlying particle distribution and/or powering
mechanism. \S2 derives the main equations governing anisotropic
Compton scattering in the context of gamma-ray binaries and discusses
the principal characteristics to expect. \S3 presents the application to the
case of LS~5039. The lightcurve and spectra observed by the HESS
collaboration are reproduced by a model taking into account
the photon field anisotropy and the attenuation due to pair creation.
\S4 concludes on the origin of the VHE emission from this system.

\section{Anisotropic Compton scattering}

Quantities in the electron rest frame are primed and quantities in the
observer frame are left unprimed. The electron energy is $\gamma_{\rm e} m_e
c^2$, the energy of the incoming (stellar) photon is
$\epsilon_0$ and the outgoing photon energy is
$\epsilon_1$. These quantities are related in the electron rest frame
by the standard
\begin{equation}
\epsilon'_1=\frac{\epsilon'_0}{1+\frac{\epsilon'_0}{m_ec^2}\left(1-\cos\Theta'\right)}
\label{e}
\end{equation}
with $\Theta'$ the angle between the incoming and outgoing photons.
The incoming and outgoing photon energies are equal
$\epsilon'_1=\epsilon'_0$ in the Thomson scattering approximation
$\epsilon'_0 \ll m_ec^2$, or $\epsilon_0\ll m_ec^2 /[ \gamma_{\rm e}
(1-\beta \cos \theta_0) ]$ when expressed in the observer frame
($\theta_0$ is the photon angle with respect to the electron direction
of motion). Scattering is also Thomson-like
even if $\gamma_{\rm e} \epsilon_0>m_e c^2$ when the incoming and
outgoing photon have almost the same direction ($\Theta'\ll (2 m_e c^2
/ \epsilon'_0)^{1/2}$). In the observer frame there is also an angle
$\theta_{\rm crit}$ below which scattering will
be Thomson-like. This angle is defined by
\begin{equation}
\cos \theta_{\rm crit}\ga {1\over\beta}\left(1-\frac{m_e
      c^2}{\gamma_{\rm e}\epsilon_0}\right)
\label{crit}
\end{equation}
i.e. $\theta_{\rm crit}\la 60$\degr\ for $\gamma\epsilon_0=1$~MeV.
The cross-section in the electron rest frame is
\begin{equation}
\frac{d\sigma}{d\epsilon'_1d\Omega'_1}(\epsilon'_0,\epsilon'_1,\Theta')=\frac{r_e^2}{2}\left(\frac{\epsilon'_1}{\epsilon'_0}\right)^2\left(\frac{\epsilon'_1}{\epsilon'_0}+\frac{\epsilon'_0}{\epsilon'_1}-\sin^2 \Theta'\right) 
\label{g}
\end{equation}
where $r_e$ is the classical electron radius and the photon
energies $\epsilon'_{0,1}$ are related through Eq.~(\ref{e}).

\subsection{Monoenergetic beam}
It is worthwile to consider first the simple case of a monoenergetic beam of
photons scattering off a single electron. The main steps are listed
below and a detailed derivation may be found in \citet{fargion}.

In the observer frame, the incoming photon density (in
$\mathrm{sr}^{-1}\mathrm{erg}^{-1}$), normalised to the (constant)
total photon density $n_0$ (in photons~cm$^{-3}$), is
\begin{equation}
\frac{dn}{d\epsilon
  d\Omega}=\delta\left(\epsilon-\epsilon_0\right)\delta\left(\cos\theta-\cos\theta_0\right)\delta\left(\phi-\phi_0\right)
\label{i}
\end{equation}
with $\delta$ the Dirac function.
The frame origin is at the location of the
electron (the frame orientation is arbitrary). The photon density in
the electron frame is found by using the invariance of
$dn/d\epsilon$ \citep{1970RvMP...42..237B}.

The fraction of photons scattered per unit time, energy and solid angle
in the electron frame is then given by \citep{1968PhRv..167.1159J,1970RvMP...42..237B}
\begin{equation}
\frac{dN'}{dt' d\epsilon'_1 d\Omega'_1}=\iint c\frac{d\sigma}{d\epsilon'_1d\Omega'_1} \frac{dn'}{d\epsilon' d\Omega'}d\Omega' d\epsilon'
\label{m}
\end{equation}
which can be transformed to the observer frame using that the number
of photons is invariant
\begin{equation}
\frac{dN}{dt d\epsilon_1 d\Omega_1}=\frac{dN'}{dt' d\epsilon'_1
  d\Omega'_1}\frac{dt'}{dt}\frac{d\epsilon'_1}{d\epsilon_1}\frac{d\Omega'_1}{d\Omega_1}.
\label{thom}
\end{equation}
$\Omega'_1$ denotes the solid angle into which the outgoing photon is
emitted and $\cos\Theta'=\cos\theta'\cos\theta'_1+\sin\theta'\sin
\theta'_1 \cos(\phi'_1-\phi')$.  Defining the polar angles
$\theta_{0,1}$ with respect to the direction of electron motion, the
resulting differential photon spectrum is a function of $\gamma_{\rm e}$,
$\theta_0$, $\phi_0$, $\epsilon_0$, $\theta_1$, $\phi_1$ and
$\epsilon_1$. The integration gives a rather unwieldy expression that
can be found in the Appendix (Eq.~\ref{knom}).

In the Thomson regime ($\epsilon'_0\ll m_e c^2$), the outgoing photon
energy is unequivocally related to the incoming photon energy since
$\epsilon'_1=\epsilon'_0$. To each polar angle $\theta_1$ corresponds
a unique photon energy. In the general regime there is also a
dependence on the azimuth (see Appendix). Staying in the Thomson
regime, the total spectrum emitted by an electron follows from the
integration over $d\Omega'_1$ of Eq.~(\ref{m}) and is \citep{fargion}
\begin{equation}
  \frac{dN}{dt d\epsilon_1}=\frac{\pi r^2_e c}{2\beta\gamma_{\rm e}^2\epsilon_0}\left[3-\mu'^2_0+\frac{1}{\beta^2}\left(3 \mu'^2_0-1\right)\left(\frac{\epsilon_1}{\gamma_{\rm e} \epsilon'_0}-1\right)^2\right]
\label{thomson}
\end{equation}
where $\mu'_0=\cos \theta'_0$ and $\epsilon_1$ varies between
$\epsilon_0 (1-\beta \mu_0) / (1\pm\beta)$. This expression shows how
the emitted spectrum depends upon the angle $\theta_0$ between the
monochromatic point source and the direction of motion of the
electron. A more general expression is given in the Appendix
(Eq.~\ref{KN}).

\begin{figure*}
{\includegraphics{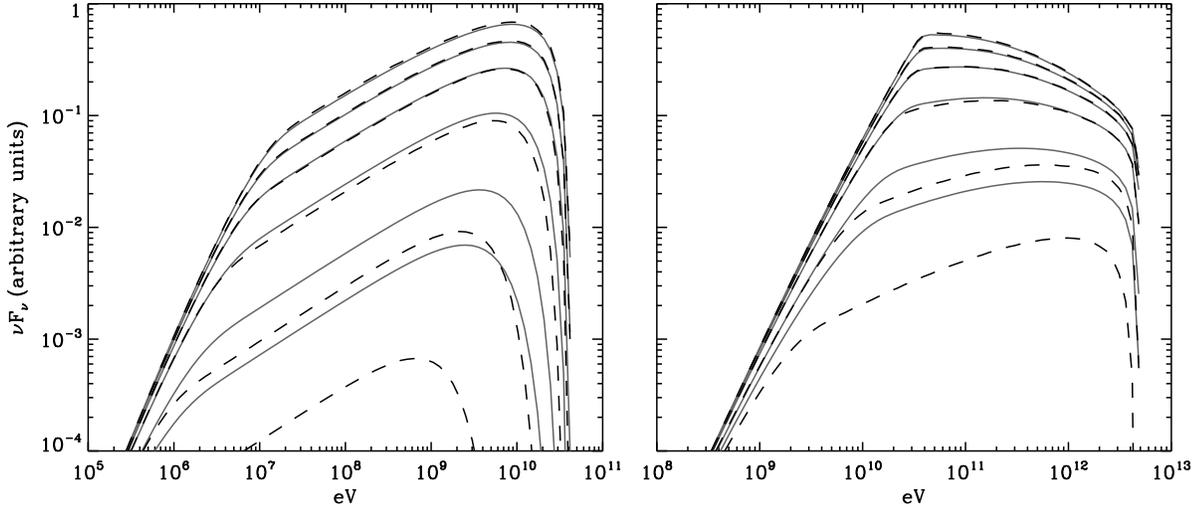}} 
\caption{Dependence of the viewing angle on the inverse Compton
  spectrum.  The source of photons is a star with $kT$=1~eV. The
  electron cloud is situated at a distance $d=2R_\star$. The electrons
  are distributed according to a power-law $dn_{\rm e}=\gamma_{\rm
    e}^{-2}d\gamma_{\rm e}$. The left panel shows the variation of the
  spectrum with angle when the interaction occurs in the Thomson
  regime (electron energy range $10^3<\gamma_{\rm e}<10^5$). In the
  right panel the interaction occurs in the Klein-Nishina regime
  (electron range $10^5<\gamma_{\rm e}<10^7$). In each panel, the
  spectrum is shown at viewing angles $\psi=$15\degr\ (bottom), 30\degr,
  60\degr, 90\degr, 120\degr\ and 180\degr\ (top).  The observer sees
  the electron cloud in front of the seed photon source when the angle
  is small. Solid lines are calculated taking into account the finite
  star size (Eq.~\ref{finite}); dashed lines correspond to the point
  source approximation (Eq.~\ref{point}).}
\label{ic}
\end{figure*}
\subsection{Kernel for spectral calculations}

The monochromatic, single photon result can be used as a kernel
to integrate over general electron and incoming photon
distributions. The total spectrum in
photons~s$^{-1}$~erg$^{-1}$~sr$^{-1}$ is then given by
\begin{equation}
  \frac{dN_{\rm tot}}{dt d\epsilon_1 d\Omega_1}=\iiiint \frac{dN}{dt
  d\epsilon_1 d\Omega_1} ~n_{0} d\Omega_0 d\epsilon_0
~\frac{dn_{{\rm e}}}{d\gamma_{\rm e} d\Omega_e} d\Omega_e d\gamma_{\rm e}
\end{equation}
where the evaluation of the kernel must take
into account the changes in electron direction with respect to the
given direction. However, this expression can be simplified.

The electron energy must be very large $\gamma_{\rm e}\gg1$ in order to emit
VHE photons. The emission is strongly forward boosted in the direction
of the electron motion by relativistic aberrations. An observer
looking at the inverse Compton emission from an isotropic cloud of
relativistic electrons sees essentially only the emission emitted by
those electrons moving within an angle $1/\gamma_{\rm e}$ from the
line-of-sight \citep[see e.g.][]{2000APh....12..335B}. Their emission is almost entirely focused into the
line-of-sight. Photons emitted slightly away from the line-of-sight
and included in the integration compensate to order $1/\gamma_{\rm e}$ for the
emission from electrons moving at larger angles. Therefore, to a good approximation, 
\begin{equation}
\int  \frac{dN}{dt
  d\epsilon_1 d\Omega_1} \frac{dn_{{\rm e}}}{d\gamma_{\rm e} d\Omega_e} d\Omega_e \approx
 \frac{dN}{dt
  d\epsilon_1}\biggr\vert_{\alpha} \frac{dn_{{\rm e}}}{d\gamma_{\rm e} d\Omega_e} \biggr\vert_{\Omega_1}
\label{el}
\end{equation}
and the spectrum will be given by
\begin{equation}
  \frac{dN_{\rm tot}}{dt d\epsilon_1 d\Omega_1}=\iiint n_{0} d\Omega_0 d\epsilon_0 \frac{dN}{dt d\epsilon_1}\biggr\vert_{\alpha} \frac{dn_{{\rm e}}}{d\gamma_{\rm e} d\Omega_e}\biggr\vert_{\Omega_1} d\gamma_{\rm e}
\end{equation}
where the kernel is given by Eq.~(\ref{thomson}) or Eq.~(\ref{KN}),
evaluated at the angle $\alpha$ between the
point-like photon source, the electron cloud and the observer. If
$\mathbf{e}_{\rm obs}$ is a unit vector from the electron cloud to the
observer and
$\mathbf{e}_{0}$ is a unit vector from the photon source to the
electron cloud, expressed using $\theta_0$ and $\phi_0$, then $\mu_{\alpha}\equiv \cos
\alpha=\mathbf{e}_{0}.\mathbf{e}_{\rm obs}$.

For scattering on an
isotropic distribution of photons, $\mathbf{e}_{\rm obs}$
can be arbitrarily oriented so that $\alpha=\theta_0$. For a
blackbody of temperature $T_\star$,
\begin{equation}
n_0
d\Omega_0=\frac{2}{h^3c^3}\frac{\epsilon^2_0}{\exp\left(\epsilon_0/kT_{\star}\right)-1}d\Omega_0\equiv
f_0 d\Omega_0
\end{equation}
For a point-like star of radius $R_\star$ at a distance $d_\star$ from
the electrons, $\mathbf{e}_{\rm obs}$ can be defined on the plane
containing the three locations so that, again, $\alpha=\theta_0$. The
photon distribution is
\begin{equation}
n_0 d\Omega_0=\pi \left(\frac{R_\star}{d_\star}\right)^2 f_0 \delta(\mu_0-\mu_\psi)\delta(\phi_0)d\Omega_0
\label{point}
\end{equation}
with $f_0$ as defined in the previous equation and where $\psi$ is the angle
between the star centre, the cloud and the observer. The integral on
$\Omega_0$ is direct so the kernel only needs be numerically integrated on
$\epsilon_0$ and $\gamma_{\rm e}$. Finally, for a star of finite size, the
integration element is
\begin{equation}
n_0 d\Omega_0=f_0 \cos \theta_0 d\Omega_0, \mathrm{~~}\phi_0 \in
[0,2\pi],~ \sin\theta_0\in [0,R_\star/d_\star]
\label{finite}
\end{equation}
and $\mu_\alpha=\cos\psi\cos\mu_0+\sin\psi
\sin\mu_0 \cos \phi_0$. This requires a quadruple numerical integral.

The electron distribution will be assumed to be isotropic in the
following so that the expression in Eq.~(\ref{el}) is a function
$f_{\rm e}$ of $\gamma_{\rm e}$ only and $\int f_{\rm e} d\gamma_{\rm e}$ gives the
total number of electrons per steradian.

\subsection{Anisotropic scattering of stellar photons\label{overall}}

Figure~1 shows example calculations of the inverse Compton spectrum
from a distribution of electrons scattering photons emitted by a star,
as seen from different viewing angles. The incoming photons have a
blackbody distribution and the electrons have a power-law distribution
$dn_{\rm e}=\gamma_{\rm e}^{-2}d\gamma_{\rm e}$. The viewing angle
$\psi$ is defined as the angle between the star, electron cloud and
observer ($\cos\psi=\mathbf{e}_\star.\mathbf{e}_\mathrm{obs}$ with
$\mathbf{e}_\star$ a unit vector from the star centre to the cloud).
Two cases are shown: one corresponding to scattering in the Thomson
regime and one for the Klein-Nishina regime. For each case, results
obtained in the point source approximation and taking into account the
finite size of the star are compared.

When scattering occurs in the Thomson regime ($\epsilon'_0\ll m_e
c^2$), the maximum energy $\gamma_{\rm e}^2 \epsilon_0
(1+\beta)(1-\beta \cos \psi)$ decreases with decreasing viewing angle
$\psi$, i.e. when the electrons move in front of the star as seen by
the observer (left panel of Fig.~\ref{ic}). This is to be expected as
the electrons are then forward scattering radiation that is less
energetic in their rest-frame than in the head-on case because of the
$1-\beta \cos\psi$ term in the Lorentz transform. The other effect is
a lower rate of emission for low $\psi$ (as can be directly deduced
from Eq.~\ref{thom} and seen in the left panel of Fig.~\ref{ic}). This
is also due to the decrease in the density of incoming photons in the
electron rest frame when both particles move in the same
direction. Scattering is more likely to occur when the particles
collide head-on \citep[e.g.][]{2000A&A...354L..53S}.

These effects are pronounced in the point source approximation and are
diluted when taking into account the finite size of the star (see
dashed lines compared to full lines in Fig.~\ref{ic}). With a star of
finite size, electrons see incoming photons from a variety of angles,
which contributes to raising the seed photon density in the electron
rest frame when $\psi=0$ (and to slightly decreasing it at
$\psi=\pi$). Because the density is tied to $1-\beta \cos \psi$, this
suggests a simple rule-of-thumb, corroborated by numerical
investigations: the effect of the finite star-size should be taken
into account when $\sin \psi\la R_\star/d_\star$ but can otherwise be
neglected. If the observer is within the cone defined by the star with
the electrons at apex, then the density of photons seen by the
electrons moving towards the observer will be significantly greater
than in the point source case. Outside of this cone, the difference
with a point source approximation is minor. In Fig.~\ref{ic}, the star
angular size seen by the electrons is 30\degr\ (defining the cone
opening angle) and the point source approximation is indeed acceptable
for $\psi>30\degr$.

When scattering occurs in the Klein-Nishina regime ($\epsilon'_0> m_e
c^2$), the maximum energy is almost constant at $\gamma_{\rm max} m_e
c^2$ regardless of viewing angle. For large viewing angles, the
spectrum is soft due to the decrease in cross-section in this regime,
just as in the isotropic case. At small viewing angles, the seed
photon energy in the rest frame of the electron is lower than in the
head-on case because of the angle dependence in the relativistic
boost, as described above for the Thomson regime. Moreover, since the
limit between Thomson and Klein-Nishina regimes is at $\epsilon_0
\gamma_{\rm e}(1-\beta\cos\psi) \approx m_e c^2$, scattering can reach
back to the Thomson regime for small enough viewing angles, regardless
of the electron energy (see Eq.~\ref{crit} in \S2). There are
two consequences. First, the amplitude of the variations with viewing
angle is smaller than in the Thomson regime, because at small $\psi$
the decrease in photon density is compensated by the larger
cross-section. Second, since there is no drop in cross-section at
small $\psi$, there can be a significant hardening of the spectrum
compared to the spectrum at larger $\psi$ (right panel of
Fig.~\ref{ic}). These spectral effects may play an important role in
modelling the emission from gamma-ray binaries, for which scattering
occurs mostly in the Klein-Nishina regime. This is investigated in the
next section.

\section{Application to LS 5039}

The influence of anisotropic scattering on the emission from gamma-ray
binaries can be sketched from the results of the previous section. If
the high energy emission is due to inverse Compton scattering off
electrons co-rotating with the binary, the viewing angle of the
observer will vary with orbital phase, inducing changes in the
observed spectrum --- all other things being set equal (particle
distribution and location, distance to the star etc).

Anisotropic scattering will most influence the emission from systems
with high inclinations, if the electrons are located in the orbital
plane. At low inclinations the changes are expected to be minor as the
scattering angle $\psi$ stays close to $\pi/2$. On the other hand, for
high inclinations the inverse Compton spectrum may change
significantly between inferior and superior conjunctions. The emission
will be intense and soft at the time of maximum attenuation by pair
production, and low and hard at the time of minimum attenuation.
Anisotropic inverse Compton emission combined with attenuation of VHE
photons can therefore play an important part in (1) reducing the
amplitude of the variations expected from a simple attenuation model;
(2) hardening the spectrum at high flux states compared to
expectations from a calculation assuming an isotropic flux.

\begin{figure}
\resizebox{\hsize}{!}{\includegraphics{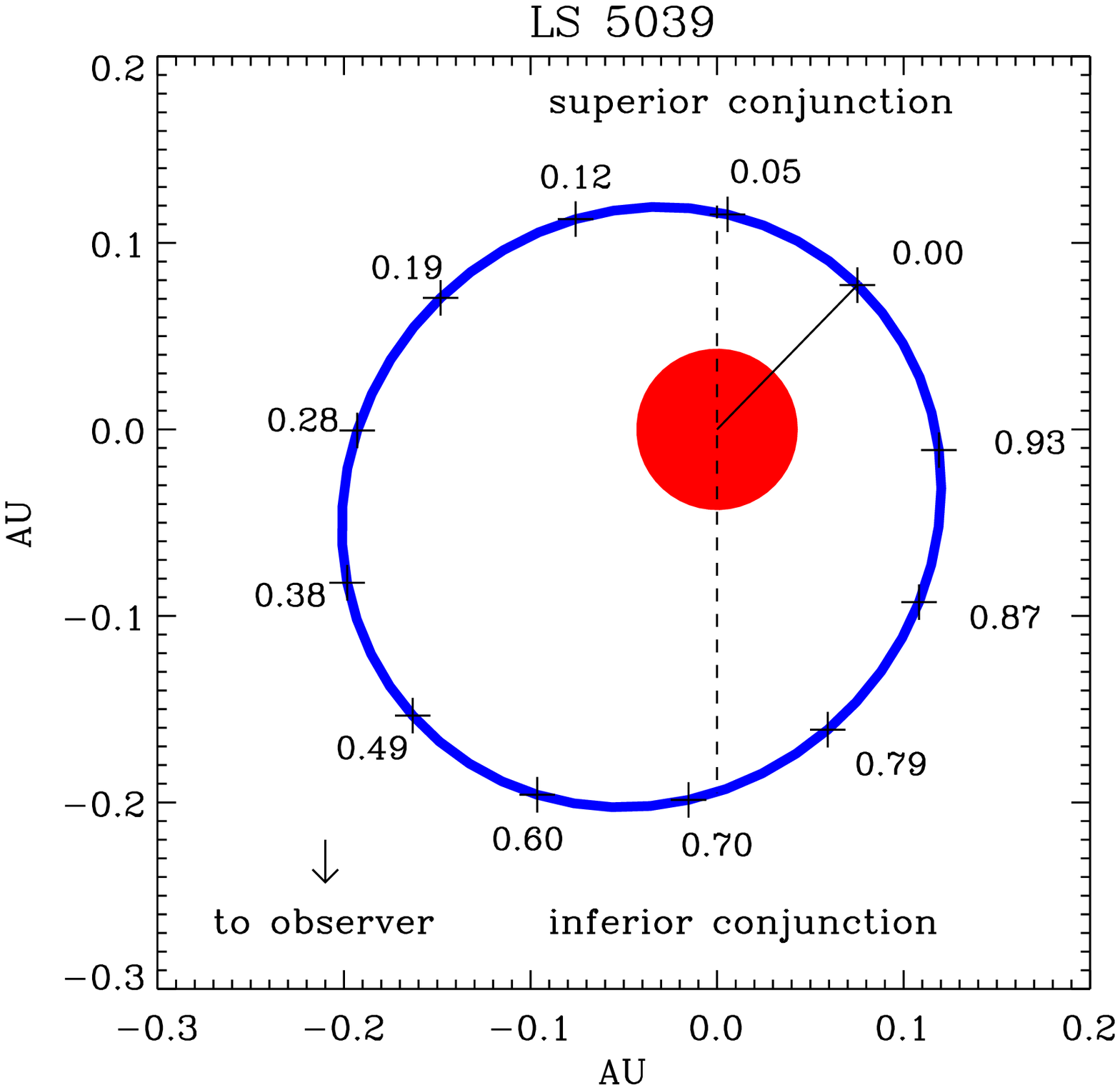}} 
\caption{The binary orbit of LS 5039 as seen from directly above. The O6.5V star radius is to scale. The binary orientation is set for an observer at the bottom of the diagram. The binary inclination on the plane of the sky is not taken into account. The numbers indicate the orbital phase (mean anomaly) at various positions. Periastron passage is indicated by a full line (orbital phase $\phi_{\rm orb}$=0). The dashed line is the line of conjunctions ($\phi_{\rm sup}\approx 0.06$, $\phi_{\rm inf}\approx 0.72$). The orbital parameters are taken from \citet{2005MNRAS.364..899C}.\label{figorbit}}
\end{figure}

\ls\ presents an ideal testbed. The massive star has an O6.5V spectrum
($T_\star=39$,000~K, $R_\star$=9.3~R$_\odot$, $M_\star$=23 M$_\odot$)
in a 3.9 day eccentric orbit ($e=0.39$) with its compact companion
\citep{2005MNRAS.364..899C}. A diagram of the binary orbit oriented on
the sky is shown in Fig.~\ref{figorbit}. The measured radial velocity
of the O star constrains the inclination to about 60\degr\ for a
neutron star companion and about 20\degr\ for a black hole. The
compact star moves from one to three stellar radii from the
  surface of the massive star.

The intensity and spectral variations have been well-established in
\ls\ by HESS observations, concluding that pure attenuation of a
constant VHE spectrum could not explain the observations to
satisfaction (see \S1). Given the above discussion, this section
examines whether taking into account anisotropic scattering provides
an improved agreement.

\subsection{The radiating electrons\label{elec}}
\begin{figure}
{\includegraphics{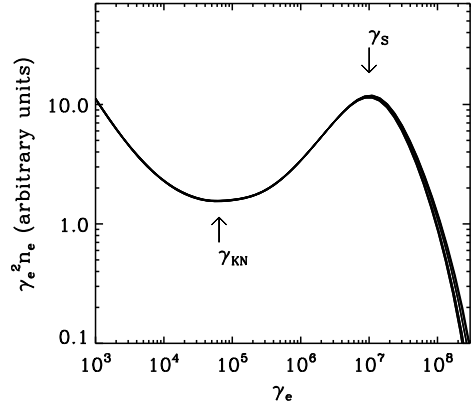}} 
\caption{Steady-state electron distribution $N_{\rm e}$ along the orbit
  of \ls. The injection spectrum is a $\gamma_{\rm e}^{-2}$ power-law with an
  exponential cutoff at $\gamma_{\rm e} \approx 10^8$ (see \S\ref{elec}). The
  magnetic field varies as $B=0.8 d_{0.1}^{-1}$~G, where
  $d_{0.1}$ is the orbital separation in units of 0.1~AU. Inverse
  Compton losses in the Thomson regime ($\gamma_{\rm e}<\gamma_{\rm KN}$) and
  synchrotron losses ($\gamma_{\rm e}>\gamma_{\rm S}$) steepen the index of
  the injected distribution by one to $N_{\rm e}\propto \gamma_{\rm e}^{-3}$.
  Inverse Compton losses in the Klein-Nishina regime dominate between
  $\gamma_{\rm KN}<\gamma_{\rm e}<\gamma_{\rm S}$, causing a hardening of the
  distribution \citep{2005MNRAS.363..954M}. The steady-state
  distribution varies little with orbital phase since $\gamma_{\rm
    S}\propto (Bd)^{-1}$ stays constant: the changes with orbital
  phase produce only a slight thickening of the line in the above
  figure.}
\label{electrons}
\end{figure}
Two main assumptions are made to calculate the emission. First, the
electrons are assumed to scatter radiation at the location of the
compact object, in a small region compared to the orbital separation.
This is a very good approximation in the pulsar wind nebula scenario
where the highest energy electrons emit the gamma-ray radiation close
to the shock. (The cooled electrons then emit in radio well away from
the system.) This may or may not be appropriate in the case of a
relativistic jet, where emission can occur at various distances along
the outflow. This is further discussed \S\ref{bh}.

Second, the adopted distribution of particles is the steady-state
distribution for constant injection of particles, taking into account
synchrotron and inverse Compton losses. The magnetic field in the
radiating zone is assumed to be homogeneous.  The radiative losses
occur on very short timescales compared to the orbital timescale so
the steady-state approximation is justified except for low energy
particles whose radiative timescale becomes longer than their escape
timescale from the radiating zone. This occurs at $\gamma_{\rm e}\approx
10^3$ (see below). The injection spectrum is a power-law
$dn_e\propto \gamma_{\rm e}^{-p}d\gamma_{\rm e}$ with an exponential cutoff at the
maximum $\gamma_{\rm max}$ allowed by comparing acceleration and
radiative timescales.

The minimum acceleration timescale for TeV electrons
($\gamma_6$=$10^6$) is set by the gyrofrequency and is $t_{\rm
  acc}\approx 0.06~ \gamma_6 / B_1$~s with $B_1$=1~G the magnetic
field intensity. The synchrotron cooling timescale is $t_{\rm
  S}\approx 770 / B_1^2 \gamma_6 {\rm ~s}$. For electrons with Lorentz
factors $\gamma_{\rm e}>\gamma_{\rm KN}\approx 6~10^4 T^{-1}_{\star,
  4}$, inverse Compton scattering of stellar photons occurs in the
Klein-Nishina regime. The corresponding timescale is $t_{\rm
  IC}\approx 20~ \gamma_6 d_{0.1}^2 / \left[\ln \gamma_6+1.4\right]
(T_{\star, 4}R_{\star, 10})^2~{\rm s}$ \citep{1970RvMP...42..237B}
with $T_{\star, 4}$=40,000~K, $R_{\star, 10}$=10~$R_\odot$ and
$d_{0.1}$ is the orbital separation in units of 0.1~AU (the LS~5039 orbital separation at periastron).

The steady-state distribution derives from a comparison of these three
timescales. Synchrotron losses dominate over inverse Compton losses
above a critical $\gamma_{\rm S}$ given by ($t_{\rm S}$=$t_{\rm IC}$):
\begin{equation}
\gamma_{\rm S}\approx 6 \cdot 10^6 ~(T_{\star, 4}R_{\star, 10})/ (B_1 d_{0.1}).
\end{equation}
At the highest energies, $\gamma_{\rm max}$ is therefore set by
synchrotron losses ($t_{\rm acc}$=$t_{\rm S}$), which gives
$\gamma_{\rm max}\approx 1.2~10^8 B_1^{-1/2} $. Assuming continuous
injection of electrons with a $\gamma_{\rm e}^{-p}$ spectrum, the steady-state
distribution is steepened by synchrotron losses between $\gamma_{\rm
  S}$ and $\gamma_{\rm max}$ to a $\gamma_{\rm e}^{-p-1}$ power-law.
Inefficient Klein-Nishina losses dominate between $\gamma_{\rm KN}$
and $\gamma_{\rm S}$, producting a hard spectrum mirroring the
decrease in energy loss rate with increasing $\gamma_{\rm e}$ in the
Klein-Nishina regime. Below $\gamma_{\rm KN}$ inverse Compton losses
in the Thomson regime result in a $\gamma_{\rm e}^{-p-1}$ power-law as in the
synchrotron case.

Steady-state distributions obtained using a full numerical calculation
follow very well the main characteristics outlined above
(Fig.~\ref{electrons}, see also \citealt{2005MNRAS.363..954M}). The
inverse Compton losses are treated in the isotropic approximation
since the magnetic field will quickly randomize particle directions.
The particles see, on average, the equivalent of an isotropic
radiation field; but the inverse Compton spectrum received by
an observer at a fixed location changes with viewing angle. In
Fig.~\ref{electrons}, the injection is a power law $\gamma_{\rm
  e}^{-p}$ with $p=2$ and the distribution between $\gamma_{\rm KN}$
and $\gamma_{\rm S}$ is roughly proportional to $\gamma_{\rm
  e}^{-1.3}$. The slope of this distribution depends on the slope of
the injected spectrum. For power-law injections $\gamma_{\rm e}^{-p}$
with hard indices ($p<2$) the slope between $\gamma_{\rm KN}$ and
$\gamma_{\rm S}$ tends to $\gamma_{\rm e}^{-1}$.  For soft indices
$p>2$, the hardening gradually disappears, reaching $\gamma_{\rm
  e}^{-2}$ between $\gamma_{\rm KN}$ and $\gamma_{\rm S}$ for an
injection with $p=3$. As discussed below, the observations of \ls\
constrain $p$ to about 2.

This steady-state distribution is a very good approximation to the
more detailed pulsar wind model of \citet{2006A&A...456..801D} for
electrons with $\gamma_{\rm e}\ga 10^3$: lower energy electrons escape from
the vicinity of the pulsar without radiating much of their energy.
More generally, this distribution should apply equally well to any
leptonic model assuming a constant injection of non-thermal particles
cooling in the vicinity of the compact object via synchrotron and inverse Compton radiation.

\subsection{Compact pulsar wind nebula: orbital lightcurve\label{orbit}}
\begin{figure}
\resizebox{\hsize}{!}{\includegraphics{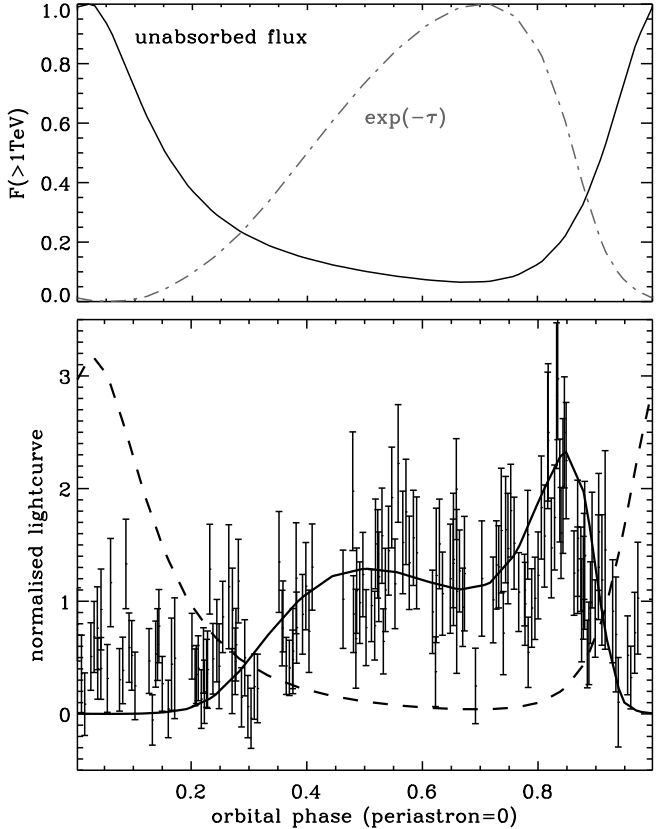}} 
\caption{Predicted orbital lightcurves for \ls\ in the case of a
  neutron star ($i=60$\degr). Top panel: the integrated photon rate
  above 1 TeV (full line) due to anisotropic inverse Compton
  scattering and the transmission $\exp(-\tau_{\gamma\gamma})$ for the
  pair production process, also integrated above 1~TeV. Inverse
  Compton scattering is minimum at inferior conjunction ($\phi_{\rm
    orb}\approx0.72$, see Fig.~\ref{figorbit}). The absorption due to
  pair production is also minimum at this time. Bottom panel: the
  resulting orbital lightcurve (full line) compared to the HESS
  observations. Combining anisotropic inverse Compton emission and
  attenuation by pair production produces a peak at $\phi_{\rm
    orb}\approx 0.8$ consistent with the observations. The agreement
  is good except at periastron where cascade emission (ignored here)
  may be important. The dashed line shows the photon rate in GLAST
  above 1~GeV (ph~cm$^{-2}$~s$^{-1}$).  The model predicts a peak in the GLAST lightcurve close to periastron and a minimum at
  inferior conjunction. The normalizations are arbitrary. The
  lightcurves are calculated using the electron distributions shown in
  Fig.~\ref{electrons}.}
\label{light}
\end{figure}
With the inverse Compton losses fixed by the geometry, the only
remaining free parameters are the slope of the injected power-law, the
total energy in radiating electrons and the value of the magnetic
field.  In the case of a compact pulsar wind nebula, the magnetic
field is determined by the conditions at the pulsar wind termination
shock.  Its intensity sets $\gamma_{\rm S}$, which in turn will
fix the frequency above which a break will be seen in the VHE
gamma-ray spectrum. In principle, $B$ may vary with orbital phase as
the eccentric orbit brings the pulsar at various radii in the stellar
wind. However, the magnetic field intensity is inversely proportional
to the shock distance from the pulsar, and the latter is roughly
proportional to the orbital separation so that $B\propto 1/d$
\citep[see e.g.][]{2006A&A...456..801D}. In this
case, the distribution of particles will not change along the orbit as
$\gamma_{\rm S}\propto (Bd)^{-1}$.

Figure \ref{light} shows the expected lightcurve at different orbital
phases with $B=0.8$~G at periastron and $p=2$ (using the electron
distribution shown in Fig.~\ref{electrons}). The orbital elements were
computed as in \citet{2006A&A...451....9D}. The unabsorbed
intensity is high close to superior conjuction and small at
inferior conjunction, as explained in \S\ref{overall}. The angle to
the observer varies between 30\degr\ and 149\degr\ whereas the angular
size of the star at the compact object is 30\degr\ at most: the finite
size of the star, taken into account in the calculation, has a
minor effect on the results. The attenuation lightcurve, computed
following \citet{2006A&A...451....9D} is also shown. It peaks at
  inferior conjunction where attenuation is minimum.

The lightcurve including both anisotropic emission and attenuation by
pair production reproduces very well the observed lightcurve. Most
notably, the combination of low attenuation, increasing photon density
and a hard inverse Compton spectrum produces a small peak after
inferior conjunction that appears to be present in the HESS
observations. The peak is a key feature of this model. This lightcurve
is very robust against changes in the value of the magnetic field
used, or even in the type of particle distribution used. At higher
inclinations, a weaker peak appears before inferior conjunction as the
variations in viewing angle cause a larger drop in inverse Compton
emission at $\phi_{\rm orb}=0.72$. However, this model still predicts
little to no flux at and after periastron because of the very strong
attenuation of the emission emitted around the pulsar. A possible
explanation is that a pair cascade develops.

The lightcurve above 1~GeV is also plotted in Fig.~\ref{light}.
Attenuation is negligible and the variations mostly follow the photon
density modulo some modifications due to the anisotropy: for instance,
the minimum is at inferior conjuction. GLAST should therefore see a
modulation in the flux from \ls\ with a peak close to periastron and a
minimum at inferior conjunction, almost anti-correlated with the HESS
modulation.

\begin{figure*}
{\includegraphics{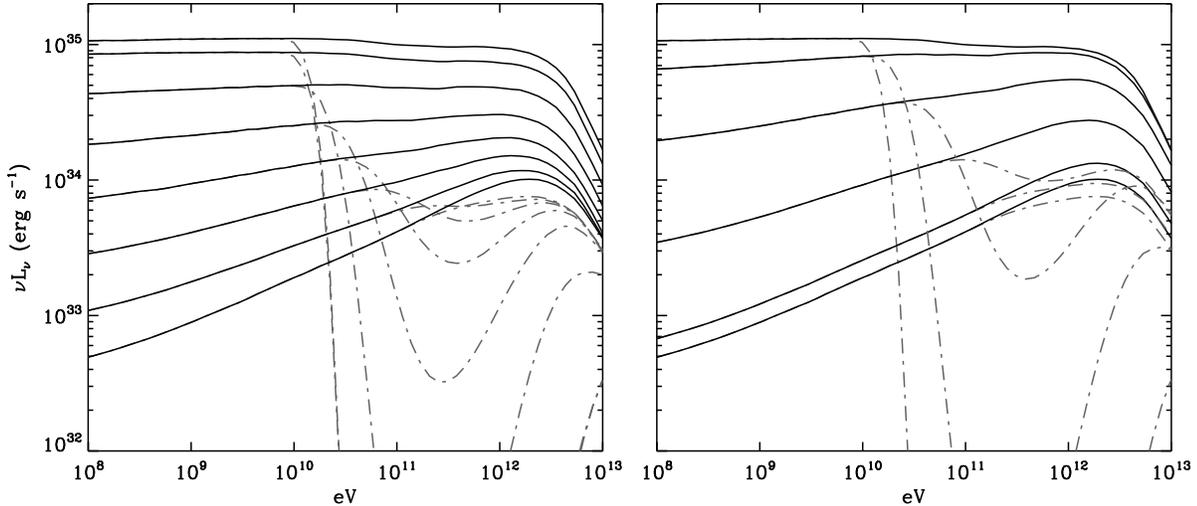}} 
\caption{Evolution of the model inverse Compton spectrum with orbital
  phase in \ls\ (neutron star case). The intrinsic emission spectra
  are shown with full lines and the dashed lines show the spectra
  after attenuation by pair production on stellar photons during the
  propagation of the gamma-rays through the system. The underlying
  electron distributions are those shown in Fig.~\ref{electrons}. Left
  panel: from top to bottom the spectra correspond to orbital phases
  $\phi_{\rm orb}$=0.03, 0.09, 0.15, 0.24, 0.34, 0.44, 0.56 and 0.66 (see Fig.~\ref{figorbit}). Right panel: the plotted orbital phases from bottom to top are 0.66, 0.76, 0.85, 0.91, 0.97 and 0.03.}
\label{sed}
\end{figure*}

A similar lightcurve has been obtained by \citet{2007A&A...464..259B},
using a complex Monte-Carlo code simulating the effects of anisotropic
scattering and the development of cascades. However,
\citet{2007A&A...464..259B} wrongly interpreted the GLAST modulation
as being due to stronger cascade emission close to periastron. As
described above, the modulation is due to a combination of increased
seed photon density and anisotropic effects and not to cascade
emission\footnote{ \citet{2007A&A...464..259B} also
  confused the phases of inferior and superior conjunctions
  (Fig.~\ref{figorbit}). The compact object is on the near side of the
  orbit (inferior conjunction) at phases 0.4-0.8 so that the broad
  maximum is not due to the stronger Compton scattering expected when
  the object is behind the star (see Fig.~\ref{sed}).  Similarly, the
  dip at phase 0.7 is not due to stronger absorption (expected at {\em
    superior} conjunction) : it actually occurs at the phase of
  minimum absorption and is due to the lower Compton emissivity at
  {\em inferior} conjunction, as described above.}.

\subsection{Compact pulsar wind nebula: phase-resolved spectra}
\begin{figure}
\resizebox{\hsize}{!}{\includegraphics{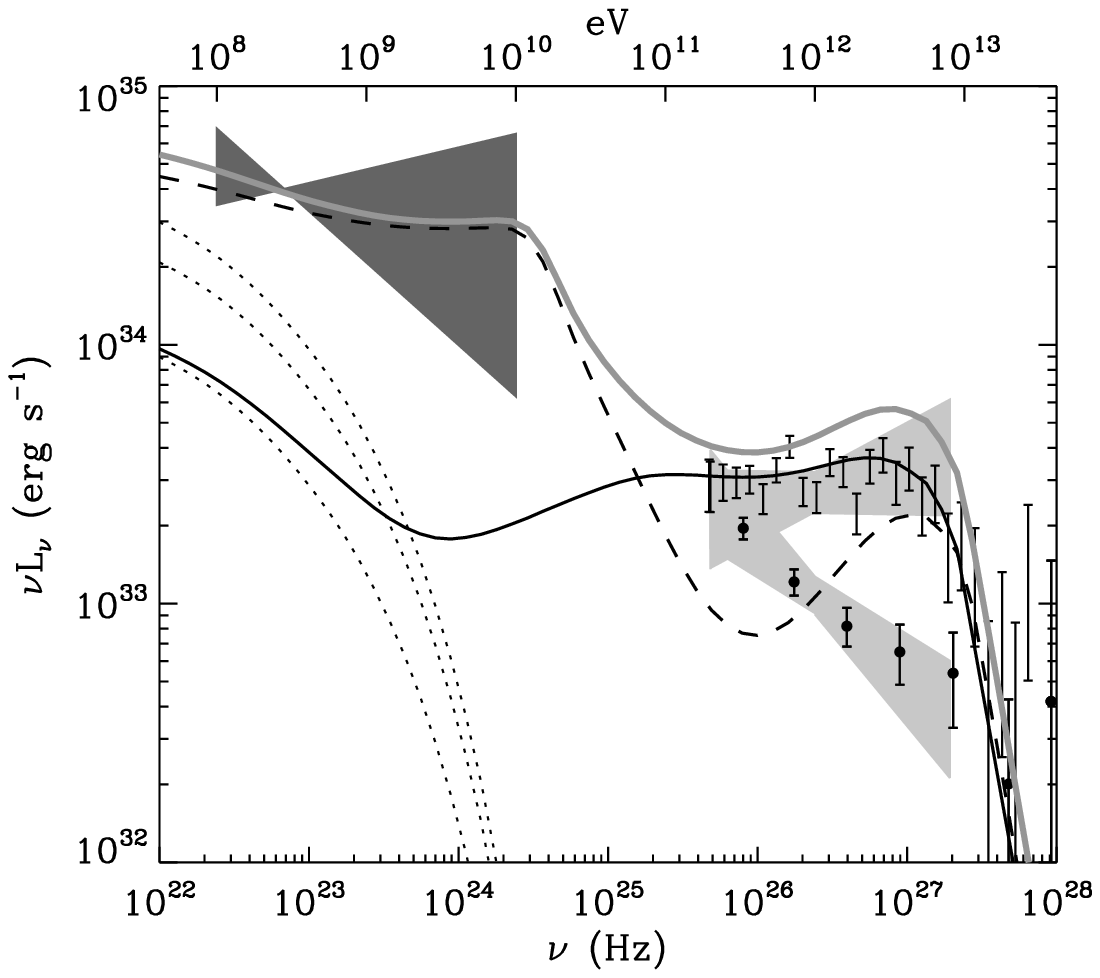}} 
\caption{Comparison with the EGRET and HESS observations of the \ls\
  model spectra for a neutron star. The EGRET bowtie is in dark grey
  and the HESS high-state and low-state bowties are in light grey
  \citep{1999ApJS..123...79H,2006A&A...460..743A}. The corresponding
  HESS deconvolved spectral points are also shown (with a dot
  identifying the low-state points).  Fluxes have been transformed to
  luminosities assuming a distance of 2.5~kpc
  \citep{2005MNRAS.364..899C}. The full grey line is the average
  spectrum calculated using the results of Fig.~\ref{sed}. It
  reproduces well the drop in flux from EGRET to HESS (the average
  HESS spectrum is close to the high-state spectrum shown). The
  high-state spectrum (full dark line) is very well reproduced
  provided the magnetic field at periastron is lower than 0.8~G. The
  low-state spectrum (dashed dark line) is not reproduced well,
  possibly because cascade emission contributes significantly at these
  orbital phases where pair production is very important or because
  the electron distribution varies along the orbit. Here, the
  synchrotron emission from the electrons is taken into account with
  $B=0.8 d_{0.1}^{-1}$~G as derived from the VHE spectrum. Its
  contribution to the spectra is shown by the dotted lines (from
  bottom to top: low-state, high-state, and orbital average). }
\label{specobs}
\end{figure}

Figure~\ref{sed} shows the evolution of the attenuated and
unattenuated spectra with orbital phase. These were used to produce
the lightcurves shown in Fig.~\ref{light}. The spectra display a
complex interplay between the varying threshold for pair production,
the high absorption it causes at superior conjunction when the inverse
Compton flux is high, and the weaker but harder inverse Compton
emission at inferior conjunction. The variations in the GeV (GLAST)
range have a very large amplitude, with a flat spectrum at the highest
intensities and a hard spectrum at low intensities. This should easily
be accessible to GLAST in the very near future
\citep{2006smqw.confE..68D}. Note that synchrotron emission
contributes significantly to the emission below a GeV and that this is
not taken into account in this lightcurve. Its impact is to soften the
spectrum and reduce the amplitude of the variations below a GeV (see
\S3.4 below and Fig.~\ref{specobs}).

The attenuated spectrum averaged over the full orbit is shown in
Fig.~\ref{specobs}. The hitherto puzzling drop between the EGRET and
HESS spectra is very well reproduced by the model without invoking a
cascade. The inverse Compton spectrum by itself underestimates the
EGRET flux by factors of a few but, taking into account the
synchrotron emission from the electrons using the adopted magnetic
field intensity ($B$=0.8~G at periastron and varying as 1/$d$), the
calculated synchrotron emission produces a very good match to both the
EGRET and HESS spectra. Note that the average HESS spectrum is not
shown for reasons of clarity in Fig.~\ref{specobs} but is close to the
'high' state spectrum (see below), with a slightly higher luminosity.

The two average spectra for the phase intervals of the HESS `high'
($0.45<\phi_{\rm orb}<0.9$) and `low' state ($\phi_{\rm orb}<0.45$ or
$\phi_{\rm orb}>0.9$) spectra are also shown in Fig.~\ref{specobs}
\citep{2006A&A...460..743A}. Reproducing the cutoff in the high-state
HESS spectrum strongly constrains the magnetic field intensity to
$\approx$ 0.8~G at periastron. A higher magnetic field moves the
cutoff to lower energies and is inconsistent with the data. A lower
$B$ moves the cutoff to higher energies and hardens the spectrum too
much. The high-state spectrum is rather sensitive to the value of $B$:
the acceptable range is only $B=0.8 \pm 0.2 d_{0.1}^{-1}$~G. Outside
of this range the fit does not go through the error bars of the HESS
data points.

The synchrotron emission contributes significantly below 1~GeV,
diluting the hardening of the spectrum around $\phi=0.7$ expected from
pure inverse Compton emission. Actually, a softening is predicted
below a few GeV. The GLAST lightcurve shown in Fig.~\ref{light} is not
noticeably changed (on a linear scale) by taking synchrotron emission
into account.  The hard electron distribution, naturally resulting
from inefficient Klein-Nishina losses here, is instrumental in
obtaining the flat spectrum in the HESS range. The range $\gamma_{\rm
  KN}< \gamma_{\rm e} < \gamma_{\rm S}$ of this hard distribution depends upon
the value of the magnetic field, but its shape is independently set by the
index $p$ of the injected power-law $\gamma_{\rm e}^{-p}$. With $p\la 1.7$ the
predicted HESS spectrum is too hard and the emission in the EGRET band
is too low.  With $p\ga 2.3$, the predicted HESS spectrum is too soft
and the EGRET emission is too large. Therefore, the slope of the
injected power-law is constrained to $p=2\pm0.3$.

Besides the magnetic field intensity and slope of injected electrons,
the other free parameter is the normalization of the electron
distribution. The fit was obtained for a total energy in electrons
from $\gamma_{\rm e}=10^3$ to $+\infty$ of 3$\cdot 10^{37}$~erg. This energy
corresponds to the injection of $10^{36}$~erg~s$^{-1}$ in particles,
assuming an escape timescale from the radiative zone of 30~s (longer
than the radiative timescale under consideration). In the pulsar wind
nebula the shocked electrons have a bulk velocity $\approx c/3$ so
that the escape timescale corresponds to a radiating zone of
$3~10^{11}$~cm, comparable to the shock size found for typical wind
parameters in \ls\ \citep{2006A&A...456..801D}. The estimated
injection energy rate is consistent with a reasonable pulsar spindown
power, such as that measured in \psrb\ \citep{1995ApJ...445L.137M}.

The low-state spectrum is responsible for most of the orbit-averaged
emission in the EGRET range, which is nicely fit by the model.
However, the HESS low-state spectrum is not satisfactory. This
spectrum corresponds to phases where the intrinsic inverse Compton
emission is both soft, as the observed spectrum, and intense. The
intrinsic emission is actually strongest at the times of highest
attenuation so that the two effects compensate somewhat. However, the
cross-section for pair production drops above a few TeV. Therefore,
the predicted phase-averaged low-state is not a pure power-law but
still shows hints of an attenuation line with a kink at high energies.
Changes in the electron distribution may also help to reduce the
discrepancy. A cutoff at a lower $\gamma_{\rm e}$ (i.e. a higher magnetic
field) than that shown in Fig.~\ref{electrons} would yield a better
agreement if it occurred at the appropriate orbital phases. However,
at this stage it appears more reasonable to investigate first the
impact of pair cascading on this spectrum, as this is required to
explain the detection at periastron.

The model contains only three parameters: the slope of the injected
power-law, the particle distribution normalization and the magnetic
field intensity at periastron (or any other arbitrary orbital phase).
The shape of the particle distribution and the associated emission
along the orbit are then unequivocally predicted. The parameters were
adjusted so as to fit the high-state HESS spectrum. That this choice
also fits very well the EGRET observations gives strong support to
this simple-minded model, even if the low-state HESS spectrum is not
reproduced to satisfaction.
 
\subsection{Black hole jet?\label{bh}}
This subsection examines how the results are changed if the compact
object is a black hole. The main effect is that consistency with
radial velocity curves require the inclination to change to 20\degr\
(4.5~M$_\odot$ black hole). The variation in viewing angle is then
reduced to the interval 70\degr --110\degr. The electrons are still
assumed to be accelerated in the vicinity of the black hole and to
reach a steady-state distribution such as the one described above.
Here, the magnetic field has a fixed value as there is no {\em a
  priori} reason for it to change with the orbital separation. This
gives a moderate change of a factor 2 in the break $\gamma_{\rm S}$ of
the electron distribution, because the orbital separation varies by
a factor 2, in contrast to the situation described in
Fig.~\ref{electrons}.

\begin{figure}[htb!]
\resizebox{\hsize}{!}{\includegraphics{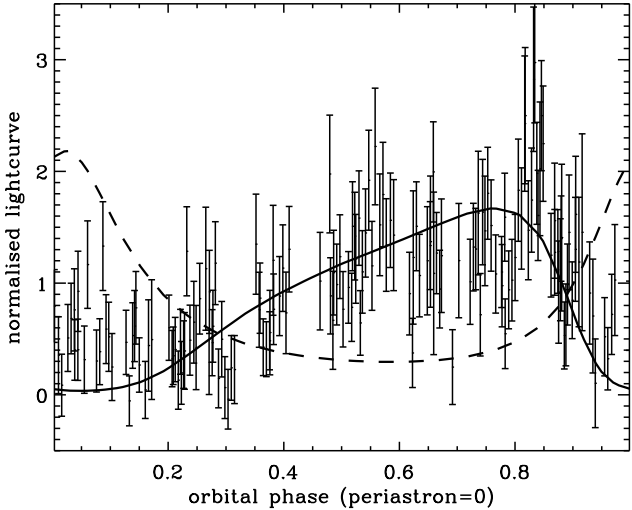}} 
\caption{Predicted orbital lightcurves for \ls\ in the case of a black
  hole ($i=20$\degr). The full line is the integrated photon flux
  above 1~TeV (HESS), the dashed line is integrated above 1~GeV
  (GLAST). The variations in viewing angle are reduced compared to the
  high inclination (neutron star) case (Fig.~\ref{light}) and there is
  only one broad maximum in the HESS lightcurve. The electron
  distribution is calculated as described in Fig.~\ref{electrons} but
  using a constant magnetic field intensity of 0.8~G.}
\label{lightbh}
\end{figure}

\begin{figure}[htb!]
\resizebox{\hsize}{!}{\includegraphics{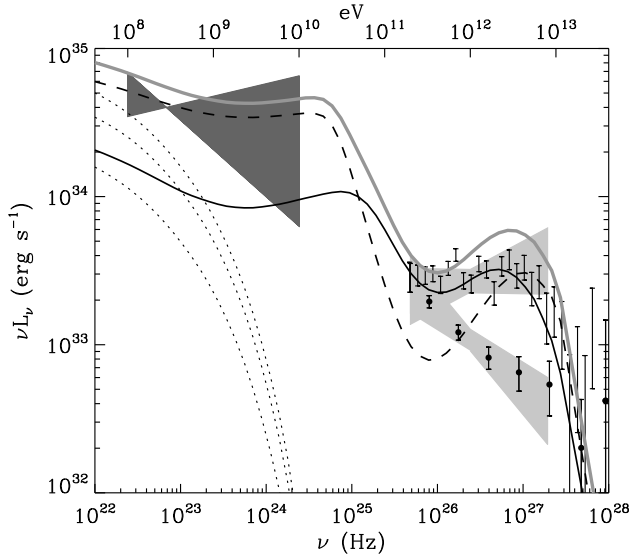}} 
\caption{Comparison with the EGRET and HESS observations of the \ls\
  model spectra for a black hole. The radiating electrons are injected
  in the immediate vicinity of the black hole. The magnetic field
  intensity used to fit the high-state spectrum is 0.8~G, constant
  throughout the orbit. The injected electrons have a power-law of
  index $p=2$. The line coding is the same as in Fig.~\ref{specobs}.}
\label{specobsbh}
\end{figure}

The orbital lightcurve and the spectra obtained with $B=0.8$~G and
$p=2$ are shown in Figs.~\ref{lightbh}-\ref{specobsbh}. In contrast
with the neutron star case, there is only one broad peak in the
predicted HESS lightcurve. This is because the reduced variation of
the viewing angle does not lead to a large drop in scattered flux at
inferior conjunction. The small peak predicted at high inclinations
(neutron star) can therefore be used as a discriminant between the two
cases.  The averaged spectra are much harder than in the neutron star
case.  The amplitude of the variation at GeV energies is less than for
a neutron star and the average flux overestimates the EGRET emission.
The poor fit of the low-state spectrum remains. Both the lightcurve
and spectra are arguably not as good fits as those obtained in the
neutron star case, but not so much as to exclude that \ls\ is seen at
a low inclination (and hence contains a black hole).

Emission from a relativistic jet may differ from the estimate above.
Any Doppler boosting will change the observed spectrum.  However, the
resolved radio emission, if interpreted as a compact jet, implies only
a moderate velocity and little boosting \citep{2000Sci...288.2340P}.
Modest Doppler (de)boosting may also be expected from the pulsar wind
emission as its post-shock speed is approximately $c$/3.  More
importantly, emission may occur all along the jet and not just be
localized near the black hole.  Far from the compact object, the
viewing angle tends to become the inclination angle ($\psi \rightarrow
i$) regardless of orbital phase\footnote{Note that two errors have
  slipped by in \citet{2006A&A...451....9D} when dealing with the case
  of a VHE source perpendicular to the orbital plane.  In the last
  equation of A.2 the angle for emission perpendicular to the plane is
  given as $\cos\psi = (d_0/d) \cos \psi_0=(d_0/d)\sin\theta \sin i$
  but should be $\cos\psi=(d_0/d) \sin\theta \sin i- (z/d) \cos i$.
  The other is that Fig.~8 (attenuation with height) was calculated at
  a fixed viewing angle of 76\degr. The conclusions are unchanged.}.
Hence, emission at progressively higher altitudes in the jet is less
and less influenced by anisotropic effects. The emission is also less
attenuated by pair production, with $\tau_{\gamma\gamma}$ negligible
at heights $\ga$~1~AU. If most of the emission occurs far in the jet,
and assuming the electron distribution stays constant, the flux 
modulation is only linked to the stellar photon density. The
result is a constant spectral shape, peak flux at periastron and a
trough at apastron. These are inconsistent with the observations.
Therefore, a jet model for \ls\ probably requires either (1) that most
of the emission occurs close to the compact object in order to
reproduce the orbital gamma-ray modulation via anisotropic scattering
and attenuation or (2) that the emission occurs away in the jet and
that some unspecified intrinsic mechanism changes the particle
distribution and/or the radiation process.


\section{Conclusion}
The anisotropic behaviour of inverse Compton scattering has a major
influence on the emission from gamma-ray binaries. In these sources,
the massive star provides a large source of seed photons with energies
around an electron-volt. If high energy electrons are accelerated in
the vicinity of the compact object, then the angle between the star,
compact object and observer changes with orbital phase. The variation
in viewing angle leads to a strong modulation in both the intensity
and spectral shape of the scattered radiation.

Scattering stellar photons to the TeV range requires very energetic
electrons with Lorentz factors $\gamma_{\rm e}\approx 10^6-10^7$. The
scattering therefore occurs in the Klein-Nishina regime. In this case,
the anisotropy results, at inferior conjunction, in a harder and
fainter spectrum than predicted using an isotropic approximation for
the incoming photons.  Crucially, inferior conjunction also
corresponds to the phase at which the produced VHE gamma-rays are less
attenuated by pair production on stellar photons.  At other phases the
emitted spectrum is close to the one obtained using the isotropic
photon field approximation and can be severely attenuated by pair
production. The result is a complex interplay that reduces the
amplitude of the variations expected from a pure attenuation model and
a hardening at inferior conjunction.

The \ls\ lightcurve and spectra were modelled using a simple-minded
leptonic model. The electrons are assumed to be accelerated
efficiently in a small zone in the vicinity of the compact object with
a standard $\gamma_{\rm e}^{-p}$ power-law. Radiative losses due to inverse
Compton emission and synchrotron emission generate a distinctive
steady-state electron distribution in this environment dominated by
stellar photons. The distribution has a prominent hardening between
the energy at which inverse Compton losses enter the Klein-Nishina
regime ($\gamma_{\rm KN}\approx 6~10^4$ in \ls) and the energy at
which synchrotron losses take over ($\gamma_{\rm S}\approx 10^7$ for a
1~G field). This is for instance the distribution found in the
vicinity of the pulsar wind shock but it applies equally well to any
leptonic model where particles are accelerated close to the
compact object. The magnetic field was allowed to vary as the inverse
of the orbital separation, as expected from a pulsar wind nebula. The
model has only three parameters: the intensity of the magnetic field, 
the normalization of the electron distribution and the slope $p$ of the
injected power-law $\gamma_{\rm e}^{-p}$.

The cutoff in the very high energy gamma-ray spectrum is very
sensitive to the magnetic field intensity, via the location of
$\gamma_{\rm S}$ in the electron distribution. Fitting the high-state
spectrum seen by HESS gives a rather constrained magnetic field
intensity at periastron of 0.8$\pm0.2$~G. This value compares well
with the values found using simple pulsar wind models which give 5
$(\dot{E}_{36} \sigma_3)^{1/2} R_{11}^{-1}$~G, where $\dot{E}_{36}$ is
the pulsar spindown power in units of 10$^{36}$~erg~s$^{1}$,
$\sigma_3$ is the ratio of magnetic to kinetic energy in the pulsar
wind in units of 10$^{-3}$ and $R_{11}$ is the distance of the shock
to the pulsar in units of 10$^{11}$~cm. Fitting the HESS high-state
spectrum also sets the injection slope to $p=2\pm0.3$, close to the
canonical value for shock acceleration. The normalization of the
electron distribution implies an injection rate of
10$^{36}$~erg~s$^{-1}$ for a radiative zone of 3~$10^{11}$~cm. These
results are remarkably consistent with the expectations for a pulsar
wind model.

The spectrum is also found to fit extremely well the EGRET
observations, adding credence to the reliability of this simple
approach. The model predicts a strong variation in the GLAST band with
a softening from high to low flux below a GeV (where synchrotron
emission dominates the spectrum) but a hardening above a GeV (where
inverse Compton emission dominates the spectrum). The HESS low-state
spectrum is not explained to satisfaction. The model fits nicely the
EGRET measurements but produces too many gamma-rays at 5-10~TeV. A
possible solution is a more complex orbital phase-dependence of the
electron distribution at selected phases.  Another solution is that
the low-state spectrum corresponds to phases of strong attenuation and
that emission from the created pairs contribute significantly to the
spectrum. Additional HESS observations near minimum flux would be
welcomed.

The orbital modulation of the HESS emission is easily reproduced. A
well-defined peak is predicted between phases 0.7-0.9 for which
evidence may already be seen in the data. The lightcurve at GLAST
energies is anti-correlated with the HESS lightcurve and has a peak at
periastron, where the stellar photon density is maximum, and a minimum
at inferior conjunction because of the anisotropic effects in inverse
Compton scattering. The GLAST spectrum below 1~GeV should be
influenced by the tail of the synchrotron emission from the highest
energy electrons. The peak synchrotron emission is at about $100$~MeV
for maximally accelerated electrons, regardless of magnetic field.
Hence, if this component is detected, it will provide evidence that
electrons are indeed accelerated with extreme efficiency in this
source.

Similar results for the magnetic field intensity and particle energy
are found when a lower inclination is used, i.e. implying a black
hole compact object rather than a neutron star. In this case, the
emission is thought to arise from a relativistic jet powered by
accretion onto the black hole. Within the assumptions of this work on
the particle distribution, it is difficult to argue that a significant
part of the emission occurs far along a jet since this does not
naturally reproduce neither the spectrum nor the lightcurve measured
by HESS. Most of the emission should still occur close to the compact
object. However, unlike in the case of a pulsar wind nebula, there is
no independent theoretical expectations in support of the magnetic
field intensity (certainly smaller than its equipartition value in the
accretion flow) and particle energy that are derived. 
Therefore, the pulsar wind nebula
model appears favoured independently of other possible considerations.

Despite the complexity of the phenomena involved in pulsar wind nebula
emission, it is found that the peculiar environment of a gamma-ray
binary, most prominently the enormous luminosity of the massive
companion, severely constrains the number of degrees-of-freedom in
the model. A simple model suffices to reproduce most of the
observations. The value of the magnetic field at the shock is found to
be tightly constrained by the HESS observations to 0.8$\pm$0.2~G and
the injection spectrum slope to $p=2\pm0.3$.
These results confirm that gamma-ray binaries are promising sources to
study the environment of pulsars on very small scales.

\begin{acknowledgements}
GD acknowledges support from the {\em Agence Nationale de la
  Recherche} and comments on an early draft from B. Giebels.
\end{acknowledgements}

\bibliographystyle{aa}
\bibliography{anis}

\onecolumn
\appendix
\section{Inverse Compton spectrum for a mono-energetic beam of photons}

The purpose of this Appendix is first to carry out the integration set out
in Eq.~(\ref{m}) and second to give an expression valid in the
Klein-Nishina regime for the total spectrum emitted by a single
electron scattering a mono-energetic beam of photons (Eq.~\ref{thomson}). The fraction of scattered photons per time, energy and steradian is
given by Eq.~(\ref{thom}), which can be expanded using
Eqs.~(\ref{e}-\ref{m})
\begin{equation}
\frac{dN}{dt d\epsilon_1 d\Omega_1}=\frac{r_e^2 c\left(1-\beta \mu_0\right)}{2\gamma_{\rm e} \left(1-\beta
    \mu_1\right)}\iiint
\left(\frac{\epsilon'_1}{\epsilon'}\right)^2\left(\frac{\epsilon'_1}{\epsilon'}+\frac{\epsilon'}{\epsilon'_1}-\sin^2
  {\Theta'}\right)
\delta(\epsilon'-\epsilon'_0)\delta(\mu'-\mu'_0)
\delta(\phi'-\phi'_0)
\delta\left(\epsilon'_1-\frac{\epsilon'}{1+\frac{\epsilon'}{m_ec^2}\left(1-\mu_{\Theta'}\right)}\right)d\epsilon'
d\mu' d\phi'
\end{equation}
where primed (unprimed) quantities are measured in the electron (observer) frame,
$\mu_{\Theta'}\equiv\cos\Theta'=\mu'\mu'_1+\sin\theta'\sin
\theta'_1 \cos(\phi'_1-\phi')$, $\mu'=\cos\theta'$,
$\mu_0=\cos\theta_0$, $\mu'_0=\cos\theta'_0$ etc. Re-arranging the last Dirac and performing
the three integrations yields
\begin{equation}
\frac{dN}{dt d\epsilon_1 d\Omega_1}=\frac{r_e^2 c \left(1-\beta
    \mu_0\right)}{2\gamma_{\rm e}\left(1-\beta
    \mu_1\right)}\left[1+\mu^2_{\Theta'_0}+\left(\frac{\epsilon'_1}{m_ec^2}\right)^2\frac{\left(1-\mu_{\Theta'_0}\right)^2}{1-\frac{\epsilon'_1}{m_ec^2}\left(1-\mu_{\Theta'_0}\right)}\right] \delta\left(\frac{\epsilon'_1}{1-\frac{\epsilon'_1}{mc^2}\left(1-\mu_{\Theta'_0}\right)}-\epsilon'_0\right).
\label{knom}
\end{equation}
The integration over $\Omega_1$ to obtain the full spectrum of
radiation emitted by the electron is simplified if 
$\gamma_{\rm e}\gg1$. In that case,
\begin{equation}
\mu_{\Theta'_0}=\mu'_0\mu'_1+\sin\theta'_0\sin
\theta'_1
\cos(\phi'_1-\phi'_0)=\frac{\mu_0-\beta}{1-\beta\mu_0}\frac{\mu_1-\beta}{1-\beta\mu_1}+\frac{1}{\gamma_{\rm e}^2}\frac{\sin\theta_1}{1-\beta\mu_1}\frac{\sin\theta_0}{1-\beta\mu_0}\cos(\phi_1-\phi_0)\approx
\mu'_0\mu'_1,
\end{equation}
which is equivalent to saying the outgoing photon is emitted
along the direction of electron motion when $\gamma_{\rm e}\gg1$. The last
Dirac can then be rewritten as a function of $\mu_1$:
\begin{equation}
\frac{dN}{dt d\epsilon_1 d\Omega_1}=\frac{r_e^2 c \left(1-\beta
    \mu_0\right)}{2\gamma_{\rm e}\left(1-\beta
    \mu_1\right)}\left[1+\mu^2_{\Theta'_0}+\left(\frac{\epsilon'_1}{m_ec^2}\right)^2\frac{\left(1-\mu_{\Theta'_0}\right)^2}{1-\frac{\epsilon'_1}{m_ec^2}\left(1-\mu_{\Theta'_0}\right)}\right] \frac{\left[1-\frac{\gamma_{\rm e}\epsilon_1}{m_ec^2}\left(1+\beta\mu'_0-(\beta+\mu'_0)\mu_1\right)\right]^2}{\left| \beta \gamma_{\rm e} \epsilon_1 +\frac{\epsilon_1^2}{m_ec^2}\mu'_0\right|}\delta(\mu_1-x)
\end{equation}
\begin{equation}
\mathrm{with~~~}x=\frac{1-\frac{\epsilon'_0}{\gamma_{\rm e}\epsilon_1}+\frac{\epsilon'_0}{m_ec^2}(1+\beta\mu'_0)}{\beta+\frac{\epsilon'_0}{m_e c^2}(\beta+\mu'_0)}.
\end{equation}
The integration over $\Omega_1$ is now straightforward, giving for the
total spectrum:
\begin{equation}
\frac{dN}{dt d\epsilon_1}=\frac{\pi r_e^2 c \left(1-\beta
    \mu_0\right)}{\gamma_{\rm e}\left(1-\beta x\right)}
\left[1+\left(\frac{x-\beta}{1-x\beta}\right)^2\mu^{'2}_0+
\left(\frac{\gamma_{\rm e}\epsilon_1}{m_ec^2}\right)^2
\frac{\left[1+\beta\mu'_0-(\beta+\mu'_0)x\right]^2}
{1-\frac{\gamma_{\rm e}\epsilon_1}{m_ec^2}\left[1+\beta\mu'_0-(\beta+\mu'_0)x\right]}\right] 
\frac{\left[1-\frac{\gamma_{\rm e}\epsilon_1}{m_ec^2}\left(1+\beta\mu'_0-(\beta+\mu'_0)x\right)\right]^2}{\left| \beta \gamma_{\rm e} \epsilon_1 +\frac{\epsilon_1^2}{m_ec^2}\mu'_0\right|}.
\label{KN}
\end{equation}
Relativistic kinematics gives the domain of variation of the scattered
photon energy $\epsilon_1$ in the observer frame. The maximum
$\epsilon_{+}$ and minimum $\epsilon_{-}$ energies in the spectrum are :
\begin{equation}
\epsilon_{\pm}=\frac{\left(1-\beta\mu_0\right)\epsilon_0}{1+\frac{\epsilon_0}{\gamma_{\rm e} m_e c^2}\pm\left[\beta^2+2\beta\mu_0\left(\frac{\epsilon_0}{\gamma_{\rm e} m_ec^2}\right)+\left(\frac{\epsilon_0}{\gamma_{\rm e} m_ec^2}\right)^2\right]^{1/2}}
\end{equation}
The angle dependence of the maximum energy in the Thomson regime is
$(1-\beta \mu_0)$. For high electron energies, in the Klein-Nishina
regime, the maximum photon energy is limited to $\gamma_{\rm e}
m_e c^2$ and becomes almost independent of angle.

\end{document}